\documentclass{emulateapj}
\begin{document}
\slugcomment{{\it The Astrophysical Journal, accepted}}

 \shortauthors{Sil'chenko, Moiseev, Afanasiev}
\shorttitle{Counterrotating gas in disk galaxies}

 \title{Two more disk galaxies with global gas counterrotation
\footnote{Based on the observations collected with the 6m
telescope (BTA) at the Special Astrophysical Observatory (SAO) of the
Russian Academy of Sciences (RAS).}}

\author{O. K. Sil'chenko \altaffilmark{1,2},
 A. V. Moiseev\altaffilmark{3},  and V. L. Afanasiev\altaffilmark{3,4} }

\vspace{2mm}

\begin{abstract}

We report a discovery of extended counterrotating gaseous disks in early-type
disk galaxies NGC~2551 and NGC~5631. To find them, we have undertaken complex
spectral observations including integral-field spectroscopy
for the central parts of the galaxies and long-slit deep spectroscopy to probe
the external parts. The line-of-sight velocity fields have been constructed
and compared to the photometric structure of the galaxies. As a result, we
have revealed full-size counterrotating gaseous disks, the one coplanar
to the stellar disk in NGC~2551 and the other inclined to the main stellar disk in
NGC~5631. We suggest that we observe the early stages of minor-merger events
which may be two different stages of the process of lenticular galaxy formation
in rather sparse environments.

\end{abstract}

\keywords{galaxies: kinematics and dynamics---galaxies: evolution---galaxies:
individual(NGC 2551)--- galaxies: individual (NGC 5631)}

\altaffiltext{1}{Sternberg Astronomical Institute, Moscow, 119991 Russia;  {\tt olga@sai.msu.su}}
\altaffiltext{2}{Isaac Newton Institute of Chile, Moscow Branch}
\altaffiltext{3}{Special Astrophysical Observatory, Nizhnij Arkhyz,
    369167 Russia; moisav@gmail.com}
    \altaffiltext{4}{ vafan@sao.ru}

\section{Introduction}

In principle, a significant role of mergers implied by the hierarchical
paradigm for the galaxy evolution must result in frequent visible
misalignments of rotation momentum between various stellar and gaseous
galactic subsystems. Especially it must be true for non-cluster
lenticular galaxies whose origin should be probably due to minor merger events.

However, findings of extended counterrotating gaseous disks are still rare. In
the Sa-galaxy NGC~3626 \citep{ciri,n3626mol}, in the S0 NGC~4546 \citep{n4546},
and in the Sb-galaxy NGC~7742
\citep{sau2,n7742we} all the gas counterrotates the stars. In the Sa-galaxies
NGC~3593 \citep{n3593_1,n3593_2,n3593mol}, NGC~7217 \citep{mk7217,we7217},
NGC~5719 \citep{n5719}, and NGC~4138 \citep{jore4138,n4138thak} the counterrotating
gas is already partly processed into stars, so these galaxies have
two stellar counterrotating disks one of which corotates the gas.
In NGC~4550 one can see already the full-size counterrotating stellar
disk whereas the counterrotating gas is mainly exhausted
\citep{n4550_1,n4550_2,n4550we,sau3,sau5};
the similar situation may be suspected in NGC~7331 \citep{prada,me7331}.
In the Sab-galaxy NGC~4826 \citep{n4826rubin,n4826rix,n4826b92,n4826b94}
and in the S0 NGC~1596 \citep{n1596} the outer gas counterrotates
the inner parts of the galaxies, certainly being accreted quite recently.
These few examples include almost all known extended counterrotating
subsystems. Statistical estimates by \citet{stat_kan} and \citet{i_stat}
put an upper limit of 8\% --12\%\ of all spiral, S0/a-Scd, galaxies to possess
such structures. For S0 galaxies the appearance of counterrotating gas may be
more frequent, \citet{km96} gives the estimate of $24\pm 8$\%; it can be
consistent with the idea of S0 galaxy (trans-)formation from a spiral by
minor merger: in such event some external gas with decoupled momentum
must be accreted. But in S0 galaxies the counterrotating gas is observed to
be mostly confined to the very central part
of the galaxies as it can be seen in the sample by \citet{b92}; the extended
counterrotating gaseous disks are rare in S0s as extended gaseous disks
in general.

Another related phenomenon is inner gaseous polar disks in disk
galaxies \citep{inpoldisk}. We found them as well in S0 galaxies with generally
small amount of gas \citep{n7280we,polars0} as in spiral galaxies with normal extended
HI disks -- NGC~2841 \citep{silvb97}, NGC~7217 \citep{we7217}, NGC~7468 \citep{n7468}.
For the inner polar disk origin the most popular
hypothesis is also external gas accretion; however in some cases dynamical
simulations predict strongly inclined circumnuclear gaseous disks produced
by secular evolution processes. The simulations by \citet{sec1} of the isolated
stellar-gaseous disk evolution gave such `polar disks' as a result of gas
redistribution in the global disk of a galaxy, if initially all the gas in the disk
counterrotated the stars. Interestingly, the old question, what is the primary,
a hen or an egg, is still actual concerning this problem. \citet{vaks82}
obtained the similar configuration, the inner polar disk plus the outer
counterrotating gas, starting from a single inner polar disk: in a tumbling
triaxial potential the outer parts of the gaseous polar disk warped in their
model so that the outer gas counterrotated the stars almost in the main
symmetry plane. By paying attention to the outer extension of the inner
gaseous polar disks found by us, we have revealed indeed some cases of the
configuration required, the inner polar disk plus the more outer counterrotation:
these are the lenticular galaxies NGC~7280 and NGC~7332 \citep{mesau5},
IC~1548 \citep{n80gr}, and the late-type spiral galaxy NGC~7625 \citep{arp212}.

An exceptionality of the extended counterrotating disks
means that there must exist some additional conditions for a disk galaxy
to retain large masses of accreted counterrotating gas; perhaps, it may be
a very low rate of the acquisition process \citep{thakryd},
or an absence of large amount of initial, `own' galactic gas in
the recipient galaxy  \citep{i_stat}. Every new disk galaxy with a
globally counterrotating gas component may in principle help to determine
these conditions. In this paper we report a discovery of two more extended
counterrotating gaseous disks in early-type disk galaxies. We present the
results of the kinematical study of the nearby S0 galaxies NGC~2551 and NGC~5631.
NGC~2551 and NGC~5631 considered in this paper are early-type disk galaxies
of intermediate luminosity. Both belong to spiral-dominated groups \citep{nog},
and both have a substantial amount of rotating neutral hydrogen with
unknown sense of rotation, according to single-dish radioobservations at 21 cm
\citep{bal_h1,h1fr}. Their main parameters retrieved in databases and from
literature are presented in the Table~\ref{tab1}.
The layout of the paper is the following. In Section~2
we describe our observations, our data reduction and some additional information.
In Section~3 we present the counterrotating gaseous disk in NGC~2551,
and in Section~4 -- the complex kinematics including inclined counterrotating
stellar-gaseous disk in NGC~5631. Section~5 contains some
discussion of the origin and possible future fate of the counterrotating gas
in these two galaxies.

\begin{table}
\caption[ ] {Global parameters of the galaxies}
\label{tab1}
\begin{flushleft}
\begin{tabular}{lcc}
\hline\noalign{\smallskip}
NGC & 2551 & 5631  \\
\hline
Type (NED$^1$) & SA(s)0/a & SA(s)$0^0$\\
$R_{25}$, kpc (LEDA$^2$) & 8.5 & 9.3\\
$B_T^0$ (RC3$^3$) & 12.78 & 12.35 \\
$M_B$ (LEDA)  & --20.0 & --20.2  \\
$(B-V)_T^0$ (RC3) & 0.92 & 0.90  \\
$V_r $ (NED), $\mbox{km} \cdot \mbox{s}^{-1}$ & 2344  &
      1979  \\
Distance, Mpc (LEDA) &  37.1 & 32.1 \\
Inclination (LEDA) & $50^{\circ}$ & $21^{\circ}$  \\
{\it PA}$_{phot}$ (LEDA) & $52^{\circ}$ & --  \\
$V_{rot} \sin i$, $\mbox{km} \cdot \mbox{s}^{-1}$
(LEDA, HI) & $83.6 \pm 5.1$ & $165.4 \pm 14.3$ \\
$M_{HI} ^4$, $10^9\,M_{\odot}$ & 1.4 & 1.4 \\
\hline
\multicolumn{3}{l}{$^1$\rule{0pt}{11pt}\footnotesize
NASA/IPAC Extragalactic Database}\\
\multicolumn{3}{l}{$^2$\rule{0pt}{11pt}\footnotesize
Lyon-Meudon Extragalactic Database}\\
\multicolumn{3}{l}{$^3$\rule{0pt}{11pt}\footnotesize
Third Reference Catalogue of Bright Galaxies}\\
\multicolumn{3}{l}{$^4$\rule{0pt}{11pt}\footnotesize
Bettoni et al.(2003)}
\end{tabular}
\end{flushleft}
\end{table}

\section{Observations and data reduction}

To study the rotation of stars and ionized gas, we use spectral data
obtained for NGC~2551 and NGC~5631 with three different spectrograph.

The Multi-Pupil Fiber Spectrograph (MPFS) of the Russian 6m telescope \citep{mpfs}
is an integral-field unit constructed following the fiber-lens principle;
due to this feature it allows to obtain panoramic spectral data over a wide
spectral range (in our case, over 1500~\AA\ with the spectral resolution
of 3~\AA ). The field of view is $16^{\prime \prime}\times 16^{\prime \prime}$,
with the sampling of $1^{\prime \prime}$ per microlens. We observed
NGC~2551 and NGC~5631 in the green spectral range containing a lot of
absorption lines and calculated the line-of-sight (LOS) stellar velocities by
cross-correlating continuum-substracted and logarithmically-binned galactic spectra
with the similarly prepaired spectra of the twilight (the Sun spectrum, of G2 spectral
type) and of G- and K-giant stars observed the same nights as the galaxies --
HD~19476 (K0III) for NGC~2551 and HD~135722 (G8III) and HD~167042 (K1III)
for NGC~5631. For NGC~2551, also the red spectral  range has been exposed to make
Gauss-fitting of the [NII]$\lambda$6583 emission line and to estimate the
LOS velocities of the ionized gas. The statistical accuracy of one-element
LOS velocity and velocity dispersion estimates with the MPFS data is about 10 km/s.

In 2007 the galaxy NGC~5631 was also observed with another integral-field
spectrograph, the SAURON
of the William Herschel Telescope at La Palma \citep{sau1}. We have retrieved these
data from the open ING Archive of the Cambridge Astronomical Data Center and have
reduced them in our manner calculating the stellar LOS velocities by cross-correlation
with the spectrum of a star observed the same night (HD~72184, K2 III, this time)
and by calculating the gas LOS velocities by measuring the baricenter positions of
the [OIII]$\lambda$5007 emission line in the continuum-subtracted spectra. The field
of view of the SAURON is $33^{\prime \prime}\times 41^{\prime \prime}$, with the
sampling of $0.94^{\prime \prime}$; the spectral resolution is 4~\AA, and the spectral
range is narrow, 4800--5350~\AA, because it is a TIGER-mode integral-field spectrograph.

After discovering central gas counterrotation in both galaxies with the
integral-field spectroscopic data, we have wanted to know a full extension of
the counterrotating gas systems. To check it, we have observed the galaxies
with the spectrograph SCORPIO of the 6m telescope in the long-slit
mode \citep{scorpio}. The slit, which length is about 6 arcmin, has
been aligned with the kinematical major axis of the central LOS velocity fields.
The red spectral range, 6100--7100~\AA, with a $2.5$\AA\, spectral resolution,
has been exposed to measure first of all the LOS velocities of the strongest
emission line in the optical spectral range, [NII]$\lambda$6583, which is free
of the underlying absorption contamination unlike
the H$\alpha$. However, the LOS velocity and stellar velocity dispersion profiles
for the stellar components have been also estimated by cross-correlating galactic spectra
binned along the slit with a template star spectrum from the library
MILES \citep{miles}. Through the library we chose spectra of
the stars HD~48433 (K1III) and HD~10380 (K3III) which provided the largest amplitudes
of the cross-correlation  function with the spectra of NGC~2551 and NGC~5631
respectively.  We  measured the errors of the stellar velocity and
velocity dispersion by using the formulae from the
classical paper by \citet{tonrydavis}. The uncertainties of the ionized-gas
kinematical parameters were estimated by Monte Carlo simulations of artificial
spectra with the noise distribution similar to that of the original data.

The log of observations is given in Table~\ref{tab2}.

\begin{table*}
\scriptsize
\caption[ ] {The observational log}
\label{tab2}
\begin{center}
\begin{tabular}{lllllrcc}
\hline\noalign{\smallskip}
Date & Galaxy & Exposure & Spectrograph & Field of view & PA(top) &
Spectral range & Seeing \\
\hline\noalign{\smallskip}
20 Aug 07 & NGC~2551 & 60 min & BTA/MPFS &
$16\arcsec \times 16\arcsec $ & $179^{\circ}$ & 4200-5600~\AA\ & $1\farcs 5$ \\
07 Mar 08 & NGC~2551 & 56 min & BTA/MPFS &
$16\arcsec \times 16\arcsec $ & $179^{\circ}$ & 5800-7200~\AA\ & $3\farcs 0$ \\
15 Jan 08 & NGC~2551 & 40 min & BTA/SCORPIO &
$1\arcsec \times 360\arcsec $ & $55^{\circ}$ & 6100-7100~\AA\ & $1\farcs 5$ \\
17 Aug 07 & NGC~5631 &  80 min & BTA/MPFS &
$16\arcsec \times 16\arcsec $ & $292^{\circ}$ & 4200-5600~\AA\ & $2\farcs 0$ \\
24 Apr 07 & NGC~5631 & 60 min & WHT/SAURON  &
$33\arcsec\times 41\arcsec$ & $307^{\circ}$ & 4800-5350~\AA\ & $1\farcs 5$ \\
06 Apr 08 & NGC~5631 & 49 min & BTA/SCORPIO &
$1\arcsec \times 360\arcsec $ & $115^{\circ}$ & 6100-7100~\AA\ & $3\farcs 5$ \\
13 May 08 & NGC~5631 & 26 min & BTA/SCORPIO &
$1\arcsec \times 360\arcsec $ & $140^{\circ}$ & 6100-7100~\AA\ & $2\farcs 5$ \\
\hline
\end{tabular}
 \end{center}
\end{table*}

\begin{figure}
\plottwo{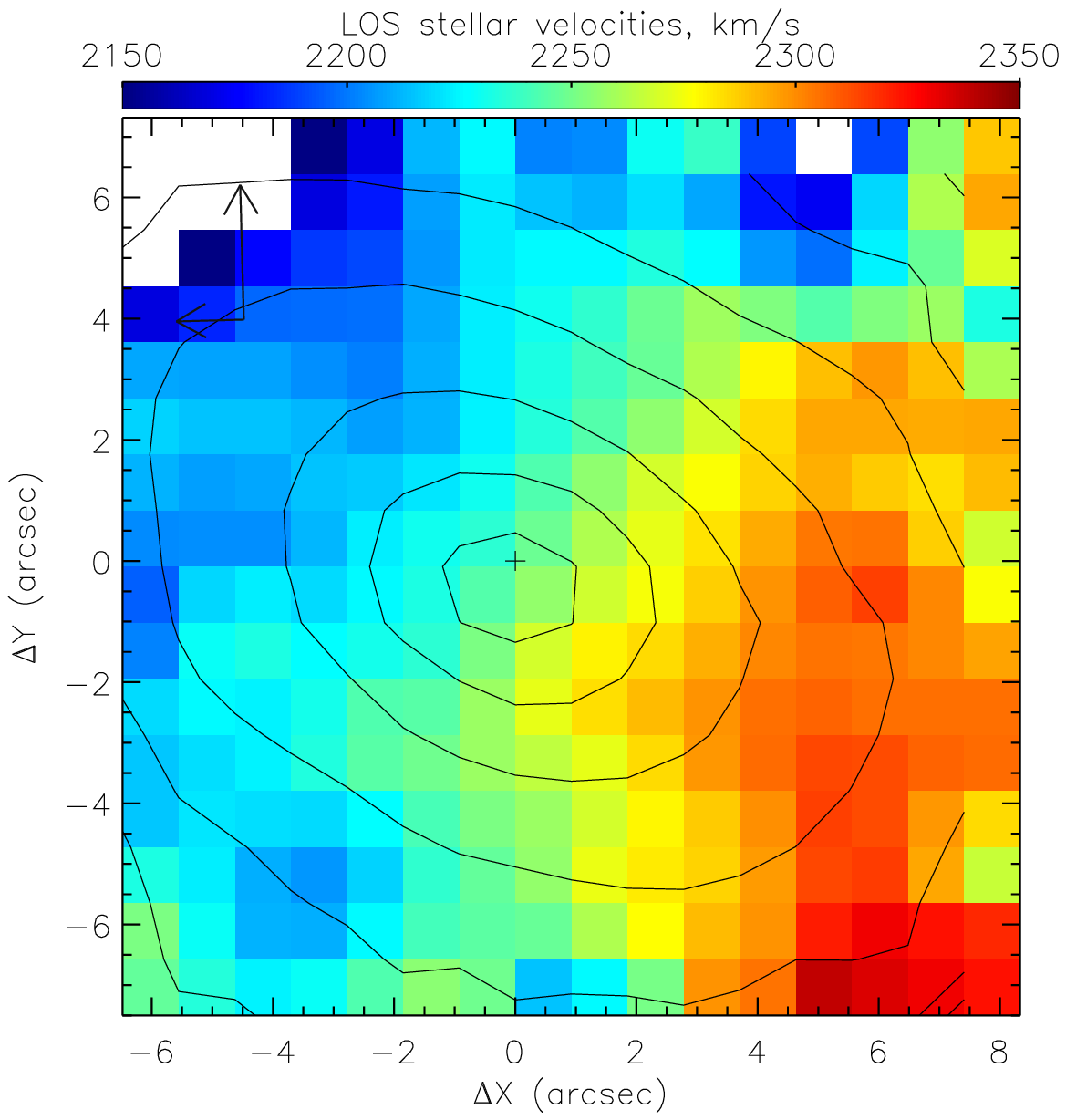}{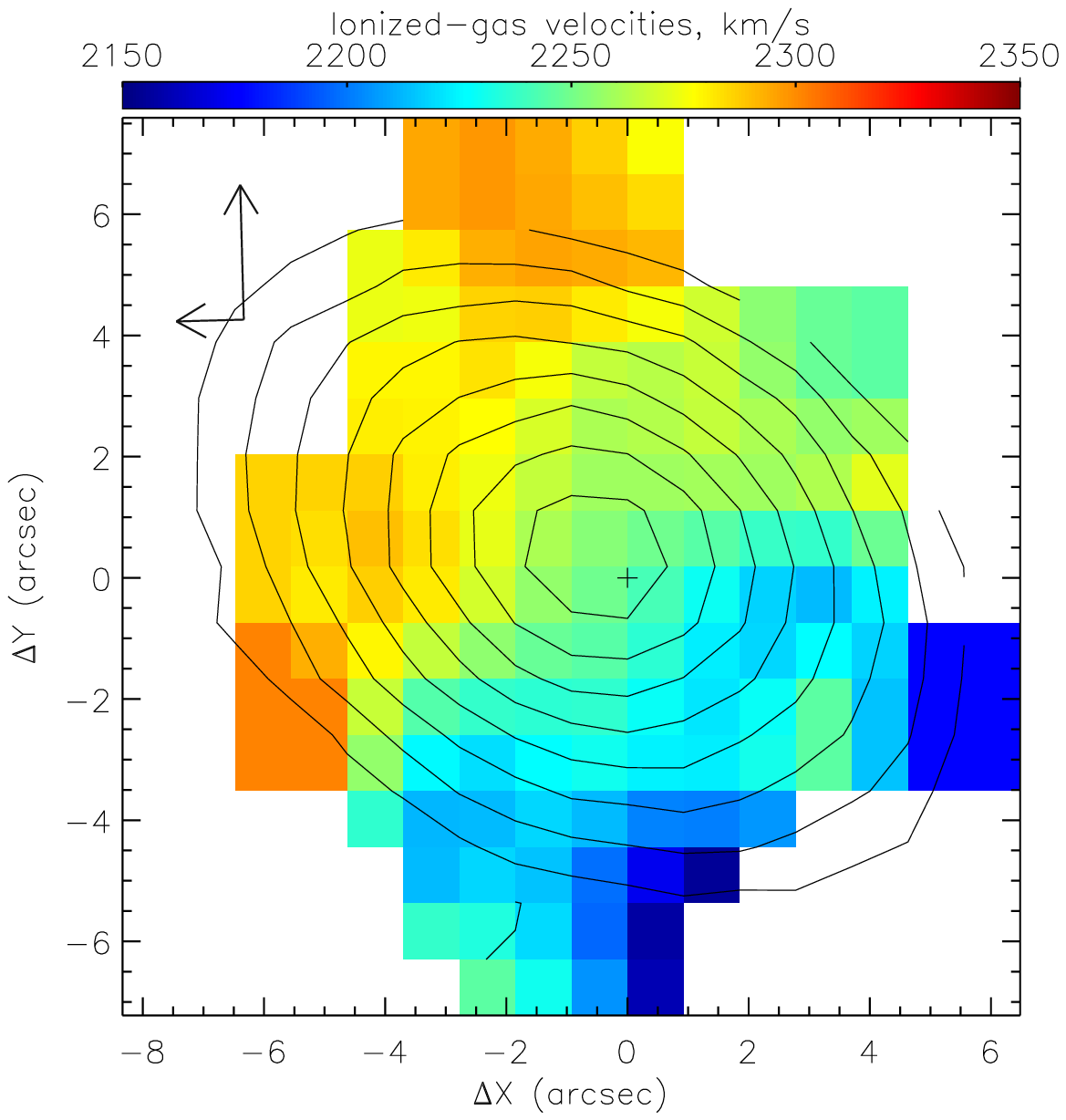}\\
\epsscale{0.5}
\plotone{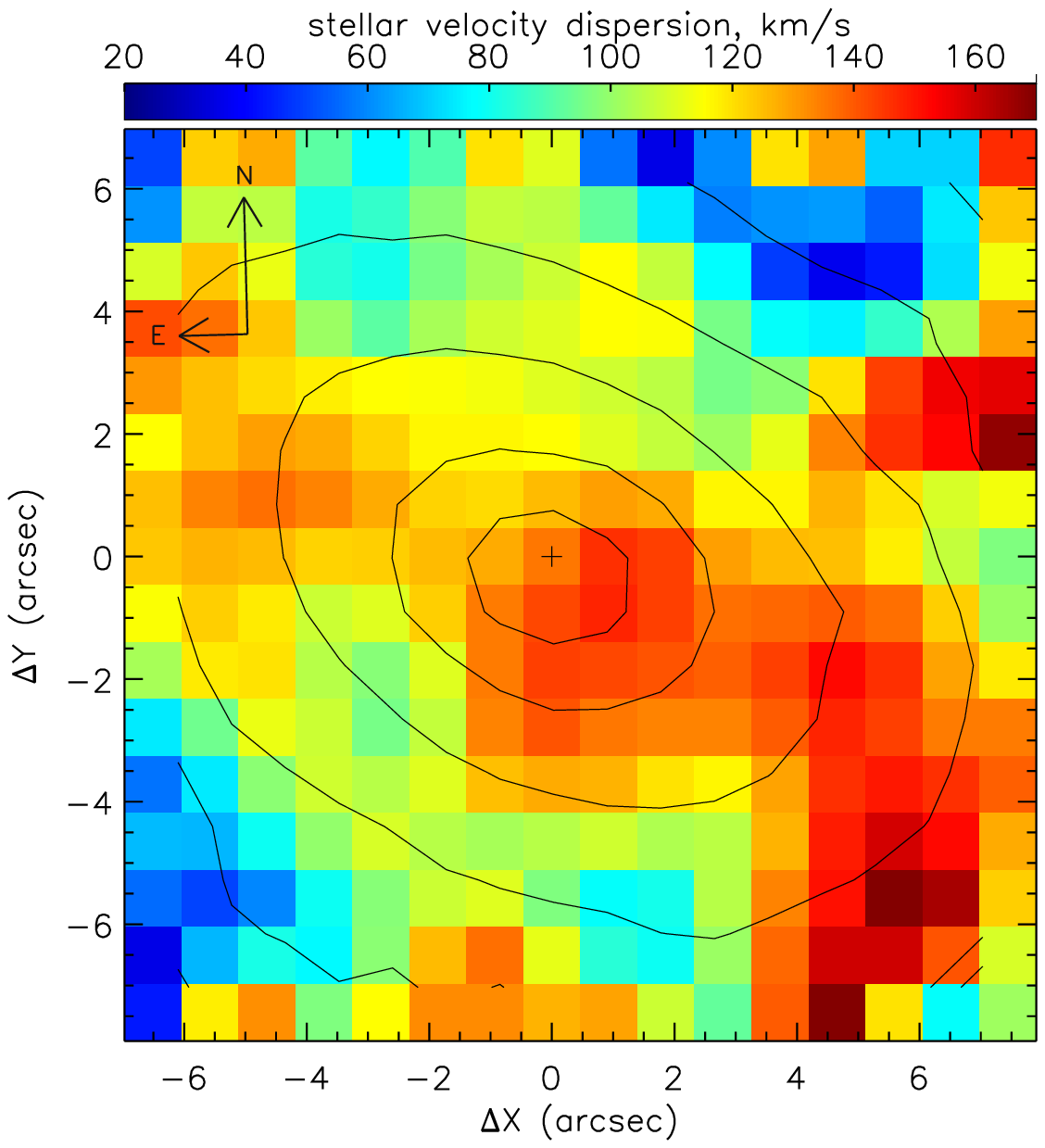}
\caption{The LOS velocity fields of stars ({\it left}) and ionized gas({\it right})
in the central part of NGC~2551 obtained with the MPFS, as well as the stellar
velocity dispersion field ({\it bottom}).
The isophotes of the continuum at $\lambda$5000~\AA\ ({\it left and bottom}) and
at $\lambda$6500~\AA\ ({\it right}) are superposed.}
\end{figure}

\begin{figure}
\epsscale{1}
\plotone{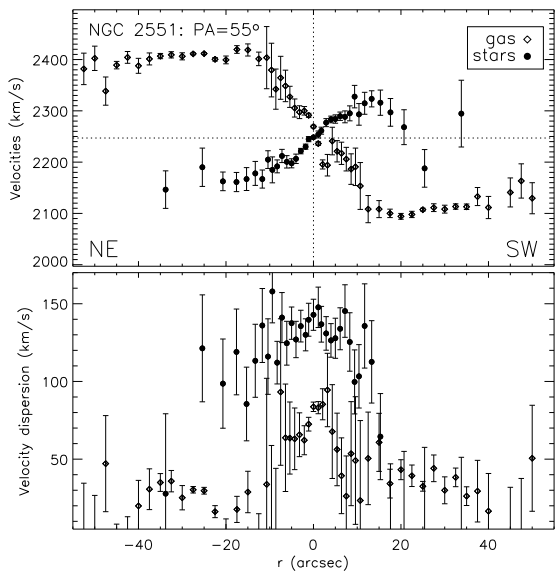}
\caption{The LOS velocity and velocity dispersion profiles of stars and ionized gas
obtained for NGC~2551 with the SCORPIO at $PA=55^{\circ}$.}
\end{figure}

\begin{figure*}
\epsscale{0.7}
\plottwo{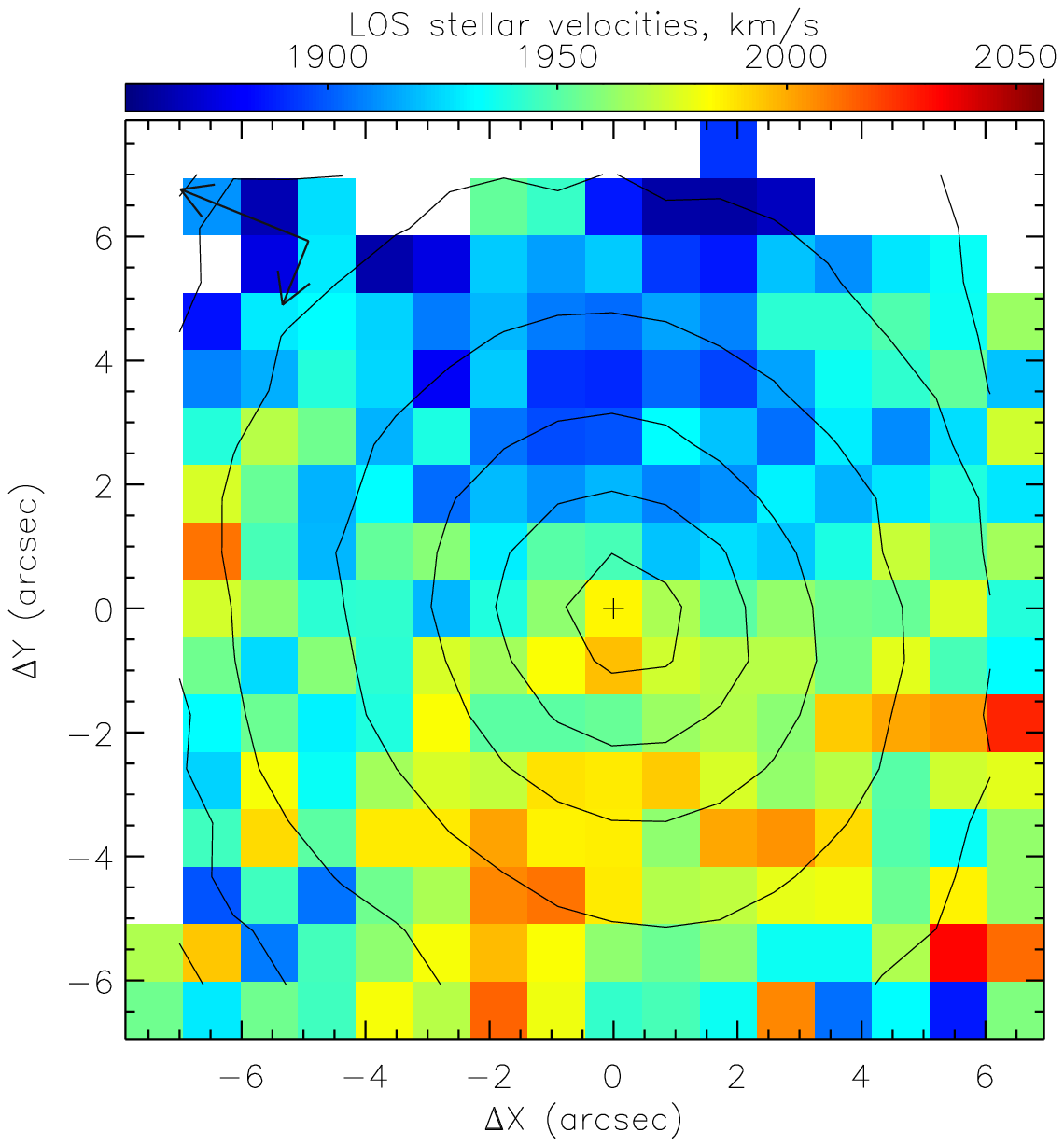}{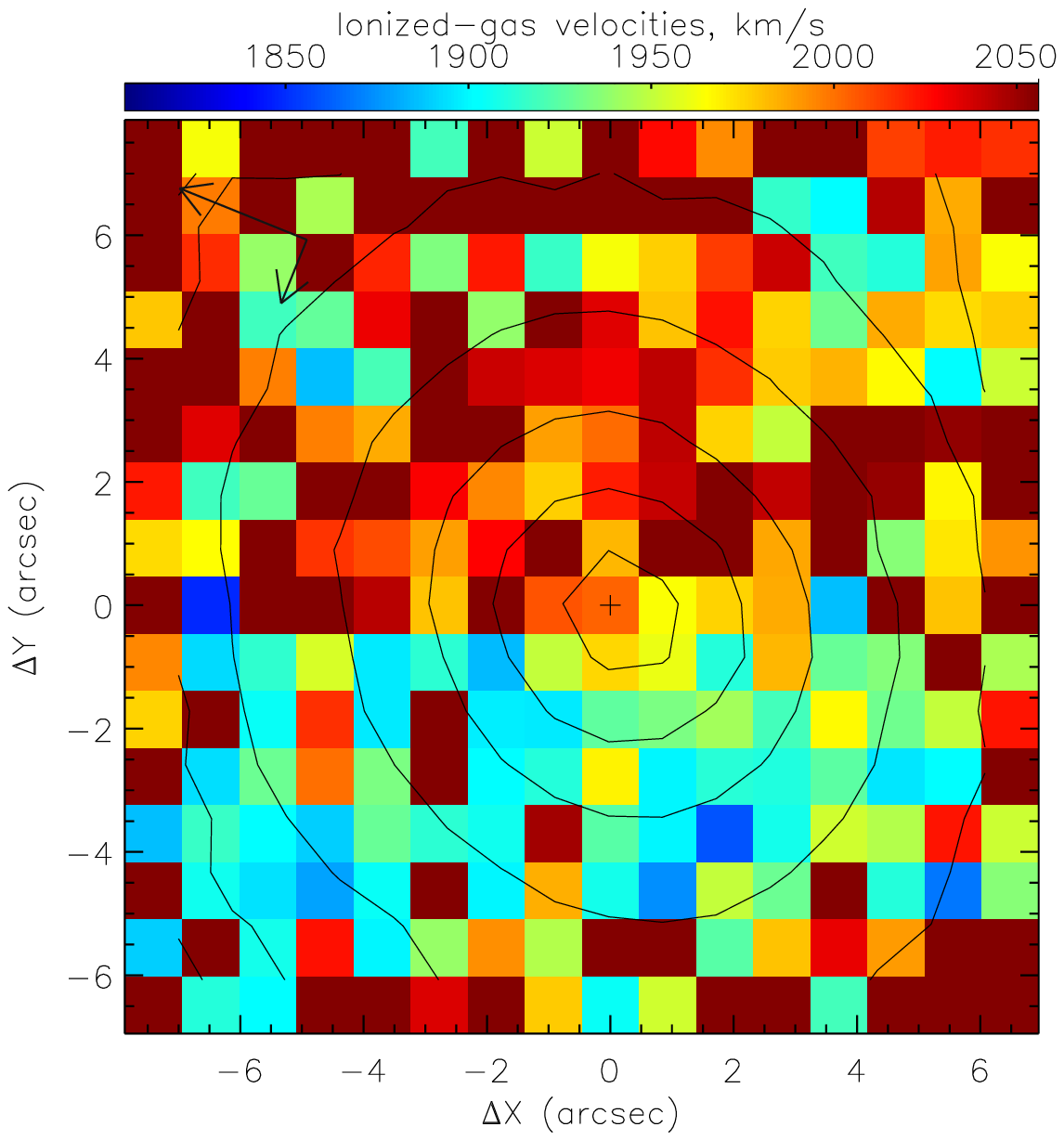}\\
\epsscale{0.7}
\plottwo{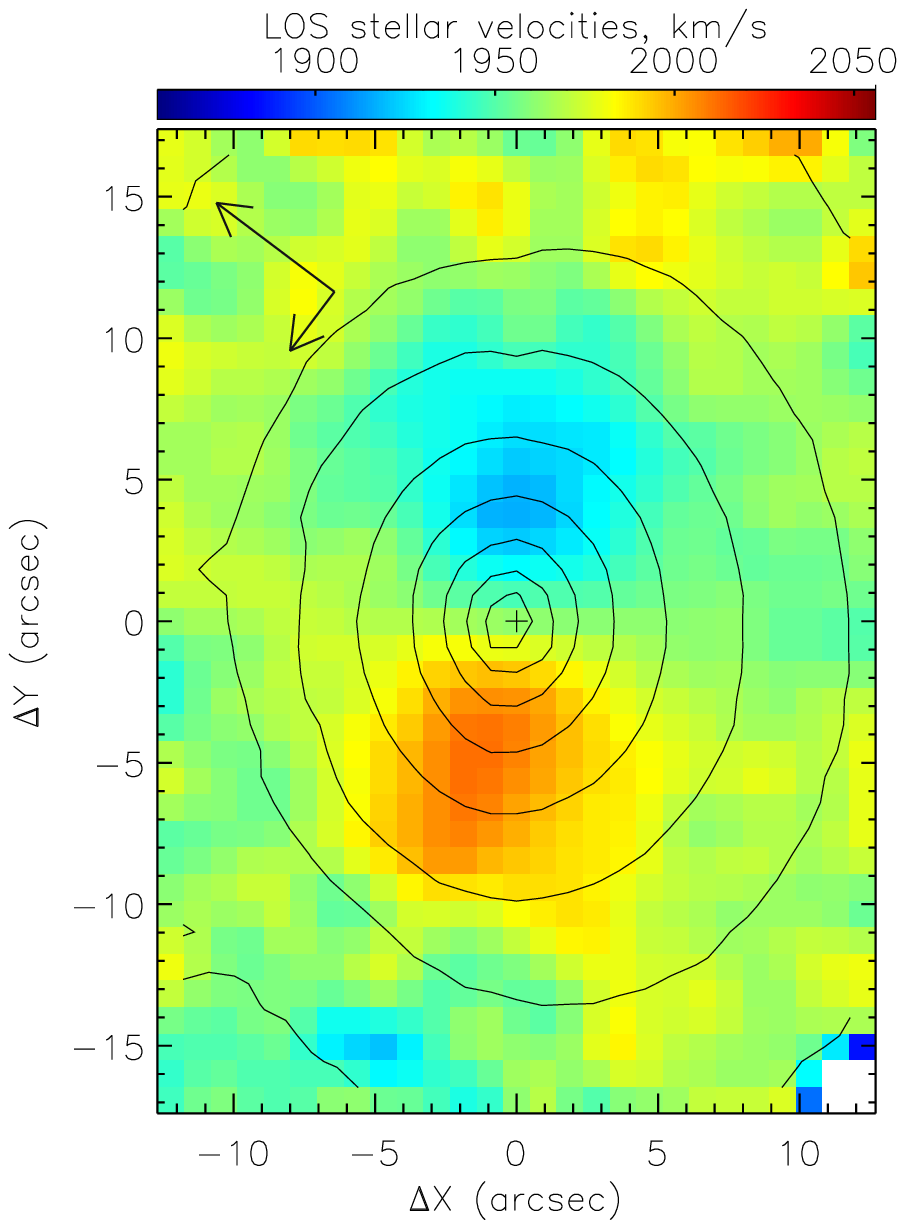}{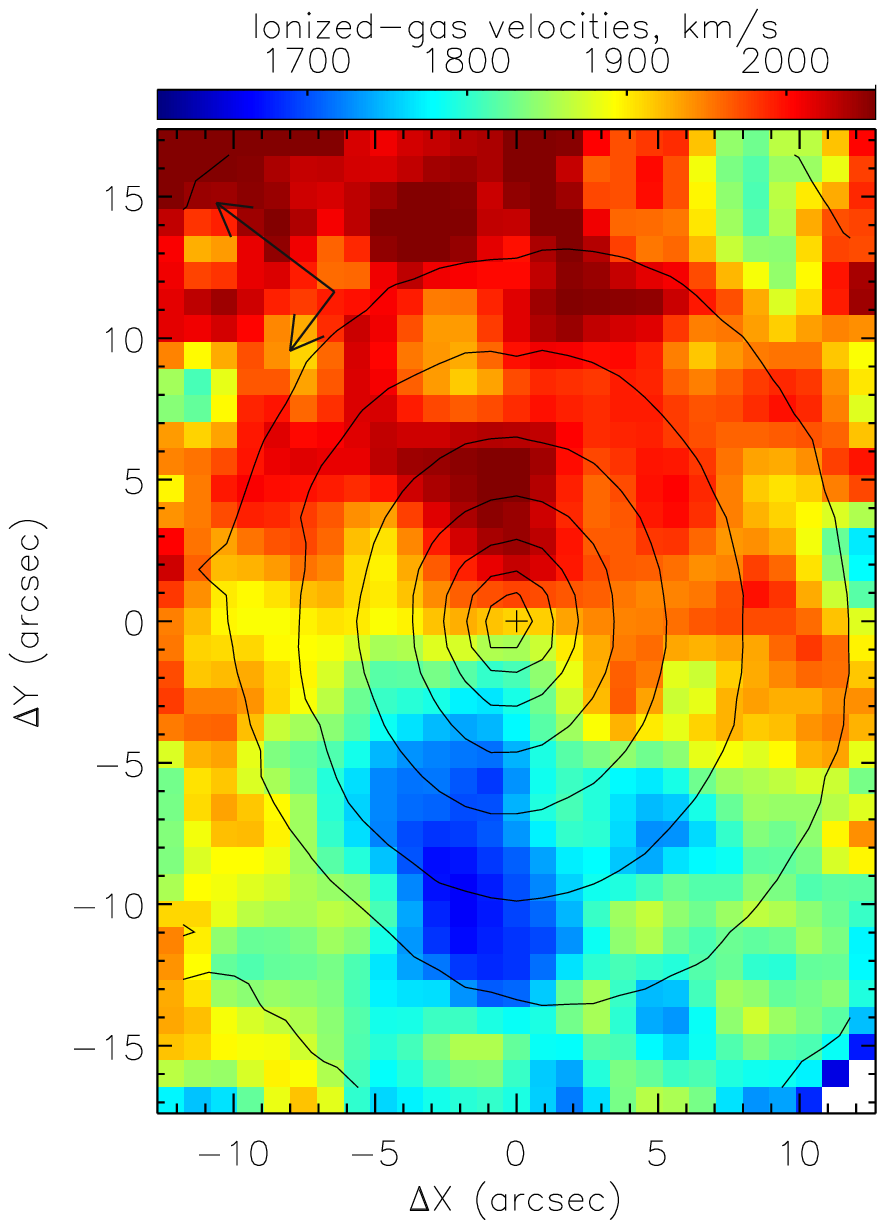}\\
\epsscale{0.35}
\plotone{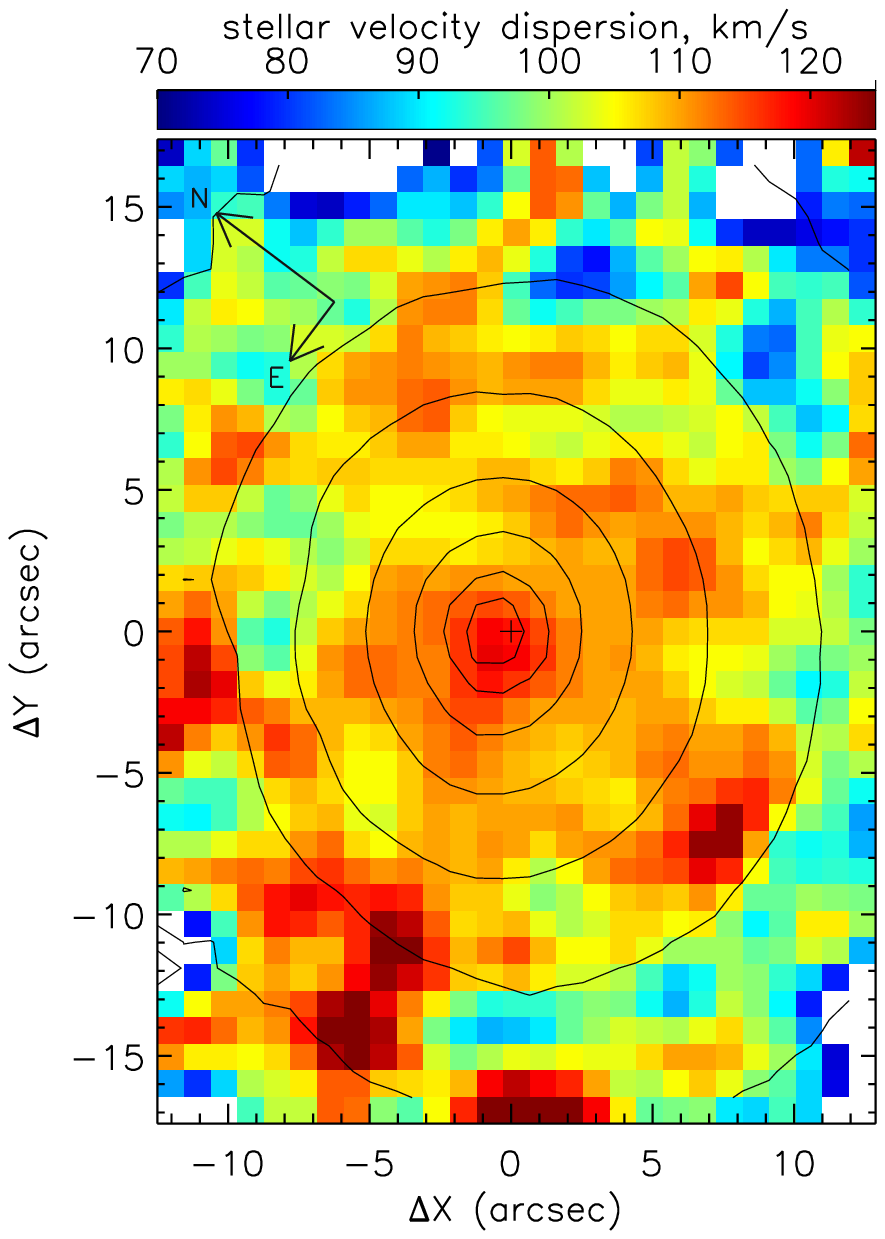}
\caption{The LOS velocity fields of stars ({\it left}) and ionized gas({\it right})
in the central part of NGC~5631 obtained with the MPFS (upper plots) and with the
SAURON (bottom plots); also the stellar
velocity dispersion field obtained with the SAURON is shown below the LOS velocity
fields. The SAURON maps are smoothed to stress
the rotation sense reverse in the outer part of the field of view.
The isophotes of the green ($\lambda$5000~\AA ) continuum are superposed.}
\end{figure*}

To analyze the morphological structure of the galaxies, we have also used surface
photometric data obtained by reducing the digital images from the HST Archive (NGC~2551, ACS/F625W) and from the SDSS/DR6 \citep{sdssdr6} data collection
(NGC~5631, $g^{\prime}r^{\prime}i^{\prime}z$-filters).

\section{Counterrotating starforming gaseous disk in NGC~2551}

Figure~1 presents the LOS velocity fields for the stellar and gaseous components in the
center of NGC~2551 which we have obtained with the MPFS. The seeing quality has been much
worse during the red-range MPFS exposure, so the visible gas rotation seems to be
slower than that of the stars; it is an effect of smearing the steep velocity gradient
by  poor spatial resolution. In general, Fig.~1 is intended only to demonstrate
gas counterrotation with respect to the stars in the center of the galaxy. The directions
of the kinematical major axis which is defined as a direction of the maximum LOS
velocity gradient are $PA_0=236^{\circ} \pm 2^{\circ}$ for the stellar component
and $PA_0=36^{\circ} \pm 10^{\circ}$ for the ionized gas. The former value coincides
with the photometrical major axis direction, $PA_{phot}=54^{\circ} \pm 1^{\circ}$
at $R=1^{\prime \prime} - 3^{\prime \prime}$ according to the HST data,
implying an axisymmetric character of the
galaxy kinematics and structure. Indeed, NGC~2551 is known to be unbarred, and its
photometric major axis direction, $PA\approx 52^{\circ} -55^{\circ}$, is constant
along the full radial extension \citep{photjap}. The fact that the isophote
ellipticity reaches its maximum, $1-b/a=0.38$, already at $R\approx 6^{\prime \prime}$
($0.1 R_{25}$) implies that the galaxy is disk-dominated. This conclusion is confirmed
by the major-axis surface brightness profile decomposition by \citet{baggett}:
according to their model, the
regular exponential disk dominates in the brightness profile of NGC~2551 starting
from $R\approx 8^{\prime \prime}$.

The next question which arose after the gas counterrotation was found with the MPFS,
was if we deal with the central decoupled gas subsystem, or the counterrotating gas is
extended over the whole galaxy. The SCORPIO gas and stellar LOS velocity profiles
obtained at $PA=55^{\circ}$ (Fig.~2) demonstrate persistence of the gas-star
counterrotation up to $R=35^{\prime \prime}$ at least. The gas rotation curve is rather
flat and extended to $R=50^{\prime \prime} \approx R_{25}$.  The projected gas
rotation velocity, 150 km/s, exceeds even the aperture HI value from \citet{bal_h1}
(Table~1) so $PA=55^{\circ}$ may be well the global disk
line-of-nodes direction. In general, the data favor coplanar stellar and gaseous
disks in NGC~2551 though rotating in opposite senses.

\vspace{15mm}

\section{Counterrotating gas in NGC~5631}

Figure~3 presents the stellar and gas LOS velocity fields for the center of
NGC~5631 constructing by using the data of the MPFS and the SAURON;
the LOS velocities of the ionized gas have been calculated by measuring the baricenter
positions of the emission line [OIII]$\lambda$5007.
The stars and the ionized gas counterrotate over the whole field of view of the MPFS, and there is a hint of the rotation reverse for the stars at the edges of the SAURON field of view.
The orientations of the kinematical major axes within $5^{\prime \prime}$ from
the center, $PA_0=118^{\circ} \pm 4^{\circ}$ for the stars and
$PA_0\approx 300^{\circ}$ for the ionized gas are consistent with each other
and with the photometric major axis orientation in the central part of the
galaxy, $PA_{phot}=126^{\circ} \pm 2^{\circ}$, implying coplanar axisymmetric rotation.

\begin{figure}
\epsscale{1}
\plotone{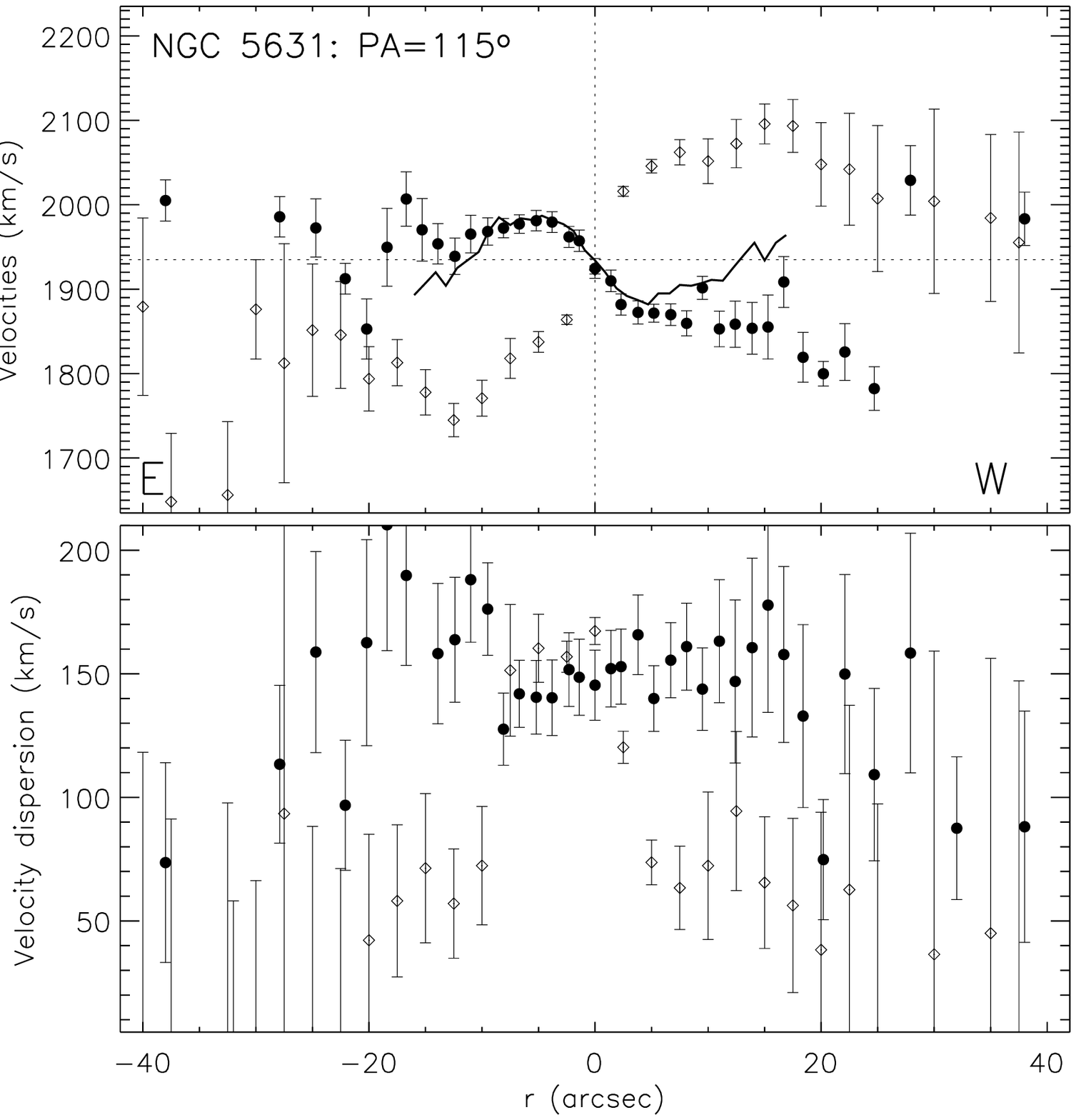}\\
\plotone{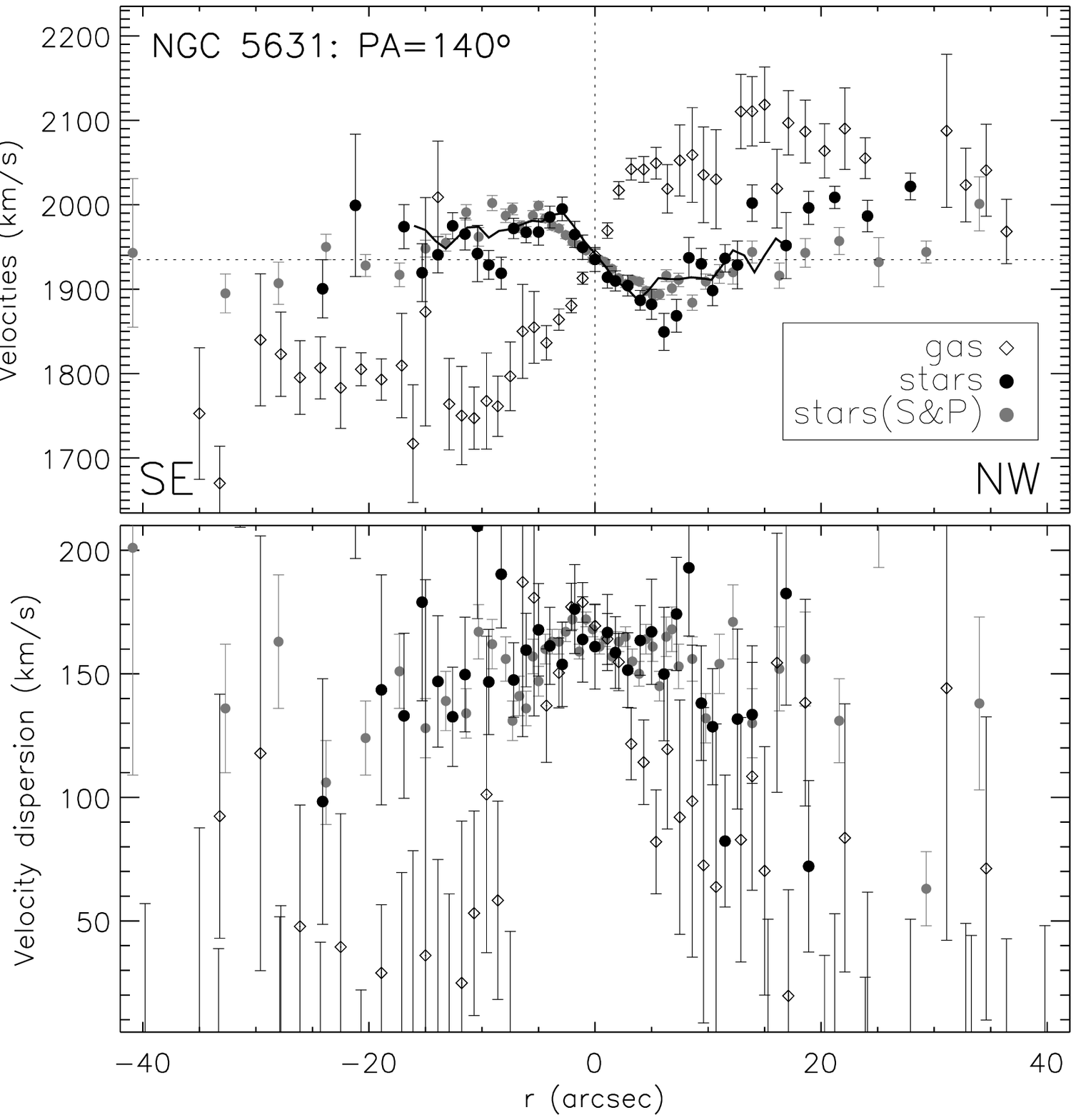}
\caption{The LOS velocity ({\it top})and velocity dispersion ({\it bottom})
profiles of stars and ionized gas obtained for NGC~5631 with the SCORPIO
in two position angles. By solid lines we superpose the digital-slit cross-sections
of the SAURON stellar LOS velocity field. Also the measurements by Simien \&\
Prugniel (2002) in $PA=135^{\circ}$ are plotted over our data for $PA=140^{\circ}$
for comparison.}
\end{figure}

\begin{figure}
\epsscale{0.9}
\plottwo{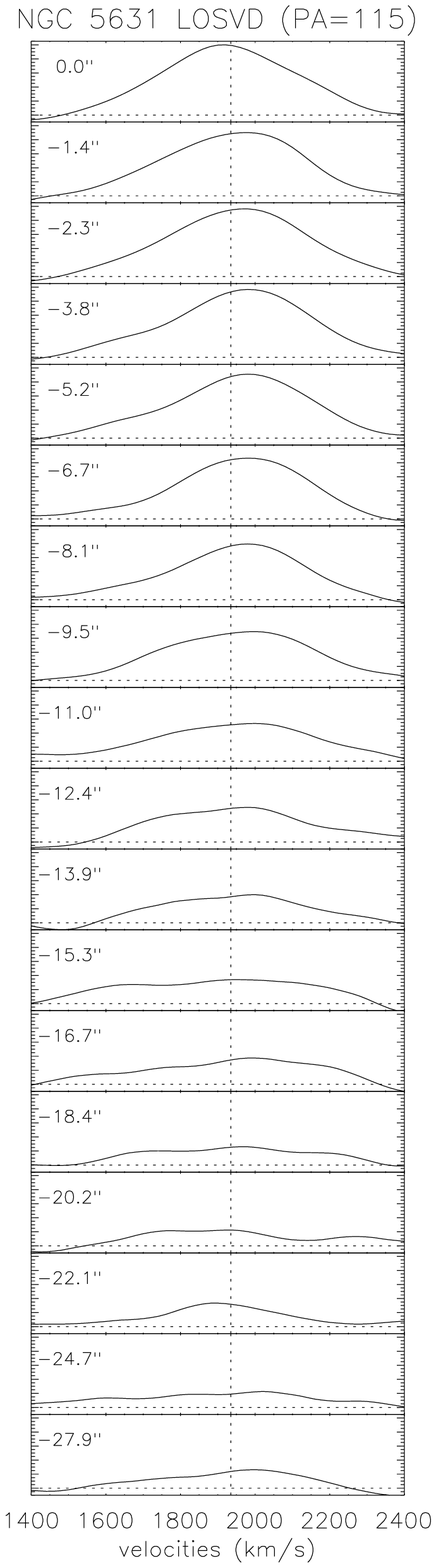}{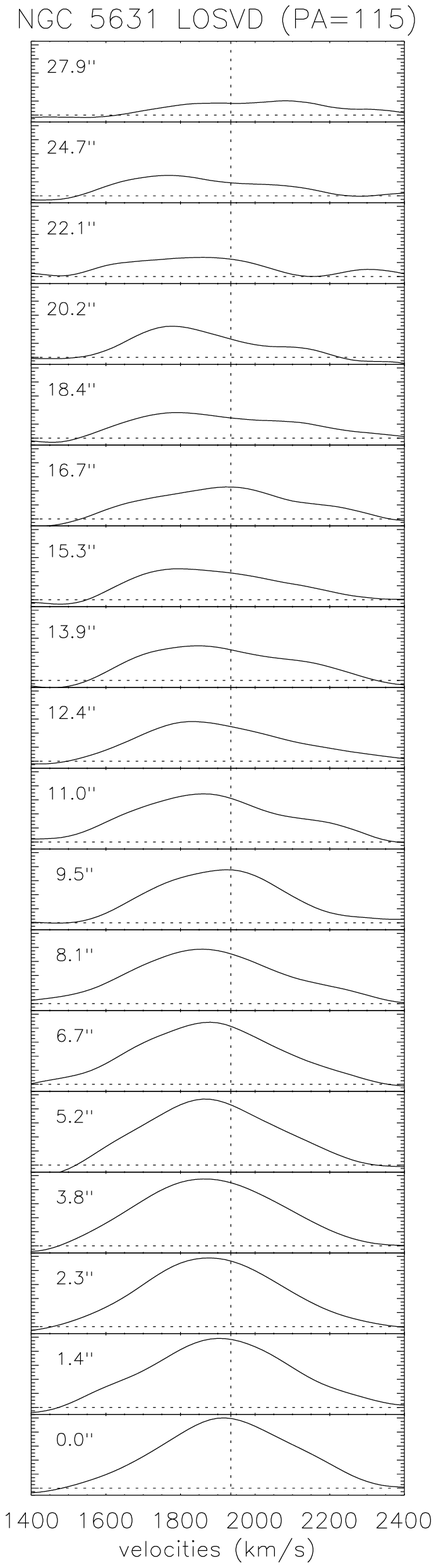}
\caption{The stellar LOSVD profile along the position angle of $115^{\circ}$
obtained for NGC~5631 with the SCORPIO. The distances from the center in
arcsec are labeled in the left corners of every plot.}
\end{figure}

The long-slit cross-sections made with the SCORPIO have shown that the gaseous
disk which rotation we observe in the center of NGC~5631 is rather extended: we see
the measurable emission lines up to $R\approx 35^{\prime \prime}$ ($0.7R_{25}$)
(Fig.~4).  The gas excitation is shock-like over the full extension of the visible
emission: the [NII]$\lambda$6583 emission line is everywhere stronger than the H$\alpha$
one. The maximum projected rotation velocity reaches about 170 km/s being
again consistent with the integrated HI data \citep{h1fr}.
The stellar component counterrotates the ionized gas up to
$R=10^{\prime \prime} - 15^{\prime \prime}$. At $R=15^{\prime \prime}$ the projected
rotation velocity of the stars falls to zero. Interestingly,  the same character of the
stellar LOS velocity profile was found by \citet{simprkin} who obtained a
long-slit cross-section at $PA=135^{\circ}$: the maximum velocity,
$v_{rot} \sin i  =51 \pm 6$ km/s, was reached at $R\approx 10^{\prime \prime}$, then
the velocity curve falled, passed through zero at  $R\approx 20^{\prime \prime}$,
and reversed its sense at $R > 20^{\prime \prime}$. The stellar velocity dispersion
profile at $PA=115^{\circ}$ demonstrates some rise at $R\sim 10^{\prime \prime}$. This
feature becomes understandable if we look directly at the LOSVD shape (Fig.~5). At
$R=-12^{\prime \prime}$ and at $R=+14^{\prime \prime}$ the LOSVD becomes asymmetric
with the hint on two peaks and remains two-peaked up to the limit of our measurements,
$R\approx 30^{\prime \prime}$. This fact results in the visible increase of the stellar
velocity dispersion and in the visible fall of the projected rotation velocity to zero:
two counterrotating stellar components compensate each other being approximated
by a single Gaussian LOSVD.

\begin{figure}
\epsscale{1}
\plotone{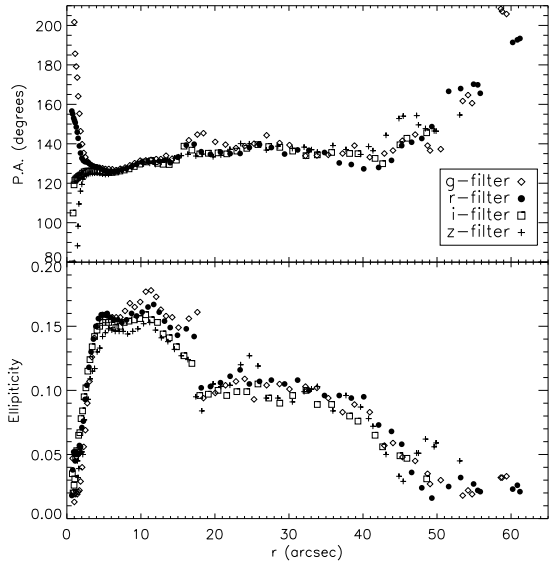}
\caption{Radial variations of the isophote parameters found for NGC~5631
from the analysis of the SDSS photometric images}
\end{figure}

\begin{figure}
\epsscale{0.9}
\plotone{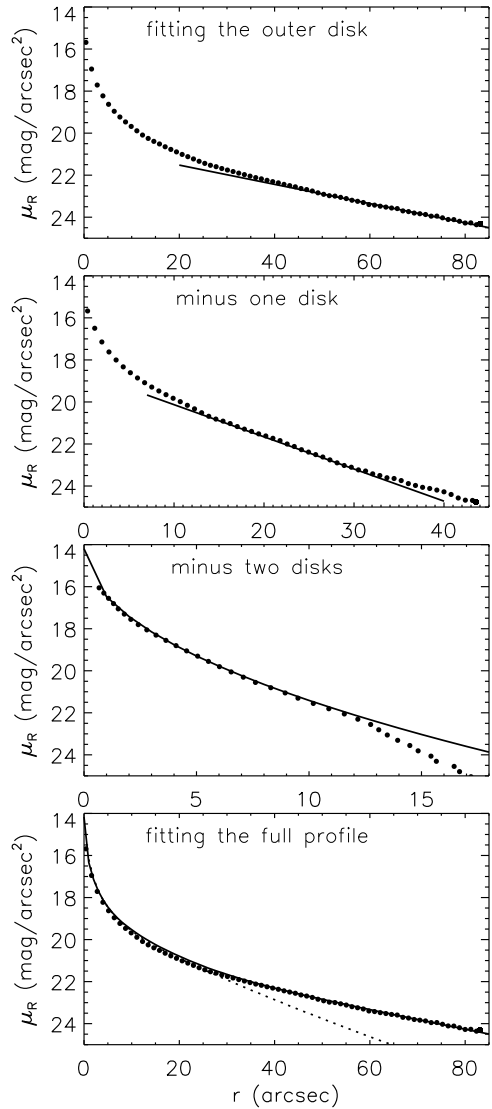}
\caption{Surface brightness profile decomposition into three
components for NGC~5631: two exponential disks with different inclinations,
$i\approx 0^{\circ}$ for the outer one and $i\approx 35^{\circ}$ for the inner one,
and one Sersic bulge with $n=2$. At the last plot we compare the observed SDSS
$r^{\prime}$-band brightness profile of NGC~5631 with our full model profile and with the
decomposition proposed by M\'endez-Abreu et al. (2008) for the 2MASS $J$-band image
of NGC~5631 (the dotted line, normalized arbitrarily).}
\end{figure}

The isophote behavior is not so simple in NGC~5631 as in NGC~2551: though the
galaxy is unbarred, the isophotes of the central part, within $R\sim 40^{\prime \prime}$
from the center, are more elliptical than those of the outer disk, and their major axis
is turning perhaps between $R\approx 3^{\prime \prime}$ and
$R\approx 25^{\prime \prime}$ from $PA\approx 120^{\circ}$ to $PA=138^{\circ}$ (Fig.~6).
The shape of the brightness profiles which we have derived from the SDSS data by averaging
the counts per arcsec over the ellipses with the parameters found by the isophote
analysis allows to divide the whole galaxy into three main components (Fig.~7).
The outer exponential stellar disk dominating at $R > 50^{\prime \prime}$ is seen
face-on: the ellipticity of the isophotes is less than 0.05, and the orientation
of the major axis cannot be determined. The inner exponential stellar component
(inner disk?) is clearly seen in the radius range of
$12^{\prime \prime} - 35^{\prime \prime}$; the orientation of the isophote major axis
is $PA=137^{\circ}-138^{\circ}$, and the isophote ellipticity after subtracting the
outer disk stays at 0.17--0.18 implying an inclination of $\sim 35^{\circ}$ under
the assumption of a rather thin disk. The surface brightness
profile of the bulge, $R < 8^{\prime \prime}$, can be approximated by a Sersic law
with $n=2$ with a high accuracy; however, the isophote ellipticity and major-axis
orientation within this central zone do not stay constant along the radius.
The recent work by \citet{m_abreu} suggests the brightness profile
decomposition of NGC~5631 into only two components, the Sersic bulge and the single
exponential disk; but as one can see in Fig.~7, their use of the shallow 2MASS
photometry does not allow them to measure the outer stellar disk in NGC~5631.

The change of the stellar rotation direction takes place within the inner disk;
the isophote ellipticity does not fall to zero at
$R = 12^{\prime \prime} - 25^{\prime \prime}$, on the contrary, it stays constant
at 0.17--0.18. So we conclude that the observed stellar LOS velocity behavior cannot
be due to a rotation plane warp in the nearly face-on galaxy, but is indeed
a manifestation of the switch of the mean stellar rotation direction inside
the zone of the photometric dominance of the inner stellar disk.

Fortunately, we can say something about the gas plane orientation too. Figure~8
presents a color map of NGC~5631 that we have constructed by using the SDSS
data; for the SDSS survey description -- see \citet{sdsstech}.
At the radius of $R=10^{\prime \prime} -15^{\prime \prime}$ one can see
a broad red (dust) ring. More exactly we see a half of the dust ring, the other half
being hidden behind the bulge. Since the gas is thought to be coupled with the dust,
we can estimate an inclination  and line-of-nodes  orientation of the gas plane at
$R=10^{\prime \prime} -15^{\prime \prime}$ under the assumption of its circular
shape. It appears to be $PA_0=122^{\circ}$
(just as the inner stellar isophotes and the kinematical major axes are directed!) and
$1-b/a=0.18$ -- just as the stellar isophotes within the inner disk.
Then the deprojected gas rotation velocity under the assumption of
$i=35^\circ$ is around 360--390 km/s within the model of circular rotation
that is high but not exceptional.

Some asymmetry of the stellar LOS velocity profiles in Fig.~4 could be then explained
if we assume that the inclined gaseous disk with the orientation parameters
deduced above contains also some stellar component, and both are coupled with
the inner stellar disk derived from the surface brightness profile decomposition.
Then the dusty stellar disk  dominating photometrically in
the radius range of $R=10^{\prime \prime} -30^{\prime \prime}$ and inclined
with respect to the face-on main stellar disk would completely hide the main
stellar component to the North of the center; just this picture is observed
in the cross-section at $PA=140^{\circ}$. The slit at $PA=115^{\circ}$
projected to the West below the line of nodes of this dusty disk, catches
the rotation of the main stellar component (a long Western receding velocity
branch, Fig.~4 {\it a}). And both cross-sections in their eastern parts demonstrate
the zero mean LOS velocities and the visible stellar velocity dispersion raising up
to 180 km/s at $R=10^{\prime \prime} -20^{\prime \prime}$ which can be treated as
a superposition of two comparable stellar components counterrotating each other.

\section{Discussion and Conclusions}

Both NGC~2551 and NGC~5631 possess the extended (up to 0.7--1.0$R_{25}$, or up to
5--7 kpc from the center) counterrotating gaseous disks. Both galaxies belong to
loose groups dominating by spiral galaxies \citep{garcia,groups_gh}.
However either NGC~2551 nor NGC 5631
have a close neighbor within the circle of 100 kpc radius, to provide interaction
and smooth gas accretion that is suggested by \citet{thakryd} to be the most
probable mechanism of massive counterrotating disk formation. The only alternative
which is available for NGC~2551 and NGC~5631 is a minor merger with a gas-rich
satellite. We do not know what morphological type the galaxies NGC~2551 and NGC~5631
had before their merging, and if they have had their own (corotating) gas.
But in any case the accreted gas had to suffer instaneous star formation triggered
by shock compression during the merging. Perhaps, in NGC~2551 and NGC~5631 we
see different stages of the same process. Ultraviolet imaging with the UIT and
later on with the GALEX has revealed an extended, up to $R=40^{\prime \prime}$,
star-forming disk in NGC~2551 \citep{uit,galex}.
An intensity ratio H$\alpha$/[NII] observed by us with the
SCORPIO implies an excitation by young stars up to $40^{\prime \prime}$; however
between  $R=40^{\prime \prime}$ and $R=50^{\prime \prime}$ we see only one emission
line, [NII]$\lambda$6583, so in this ring the gas excitation may be of shock origin.
In NGC~5631 there is no current star formation in the counterrotating gaseous disk,
and the gas is excited by shock over the full extension of the gaseous disk. Perhaps,
it is the preceding evolutionary stage with respect to NGC~2551, the accreted gas
disk has not yet settled into the symmetry plane of the main galaxy, and star formation
is only going to start in the compressed inner dusty ring at $R=10^{\prime \prime} -
15^{\prime \prime}$. And what may be the subsequent stage? Perhaps, it is NGC~4138
where a counterrotating extended gas is supplemented by the substantial
counterrotating young stellar component \citep{n4138thak} -- the evident
consequence of the star formation in the counterrotating gaseous disk.

To give a conclusion, we summarize that
by applying complex spectral methods including integral-field spectroscopy
to the central parts of the galaxies and long-slit deep spectroscopy to probe
the external parts, we have found two more global gas counterrotating systems
in non-interacting early-type disk galaxies. In NGC~2551 two counterrotating
disks, gaseous and stellar ones, may be coplanar: the orientation parameters
of the optical-band image and of the UV-band, star-formation related image
are similar. In NGC~5631 the gaseous disk
is inclined by some $35^{\circ}$ to the main stellar disk; it may contain also
significant coupled stellar component. The totality of the spectral and photometric
data give evidence for the minor merging as the most probable origin of the
counterrotating gas in these galaxies. Perhaps, we observe two different stages
of the process of lenticular galaxy formation in rather sparse group environments.

\begin{figure}
\epsscale{1}
\plotone{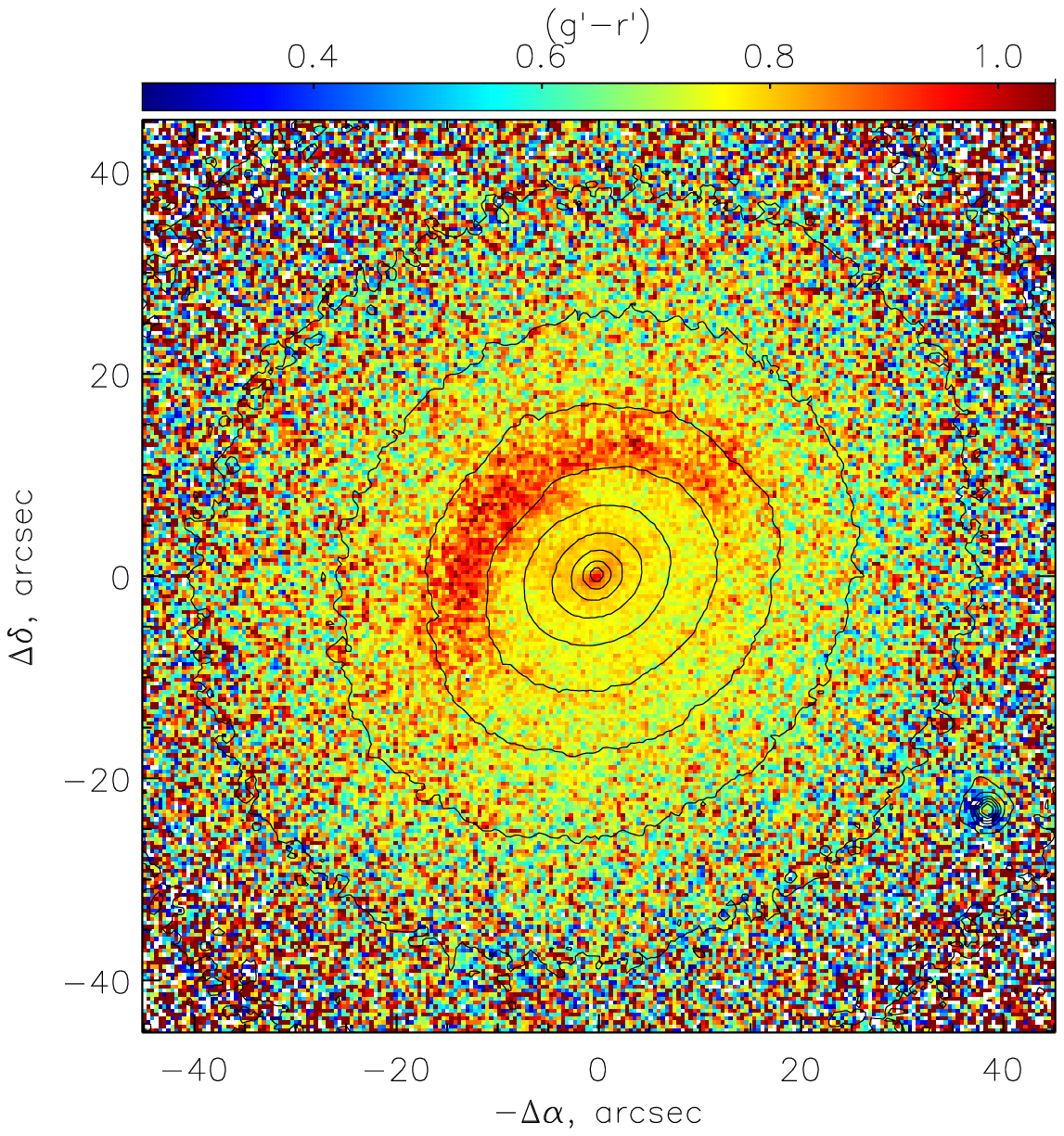}\\
\plotone{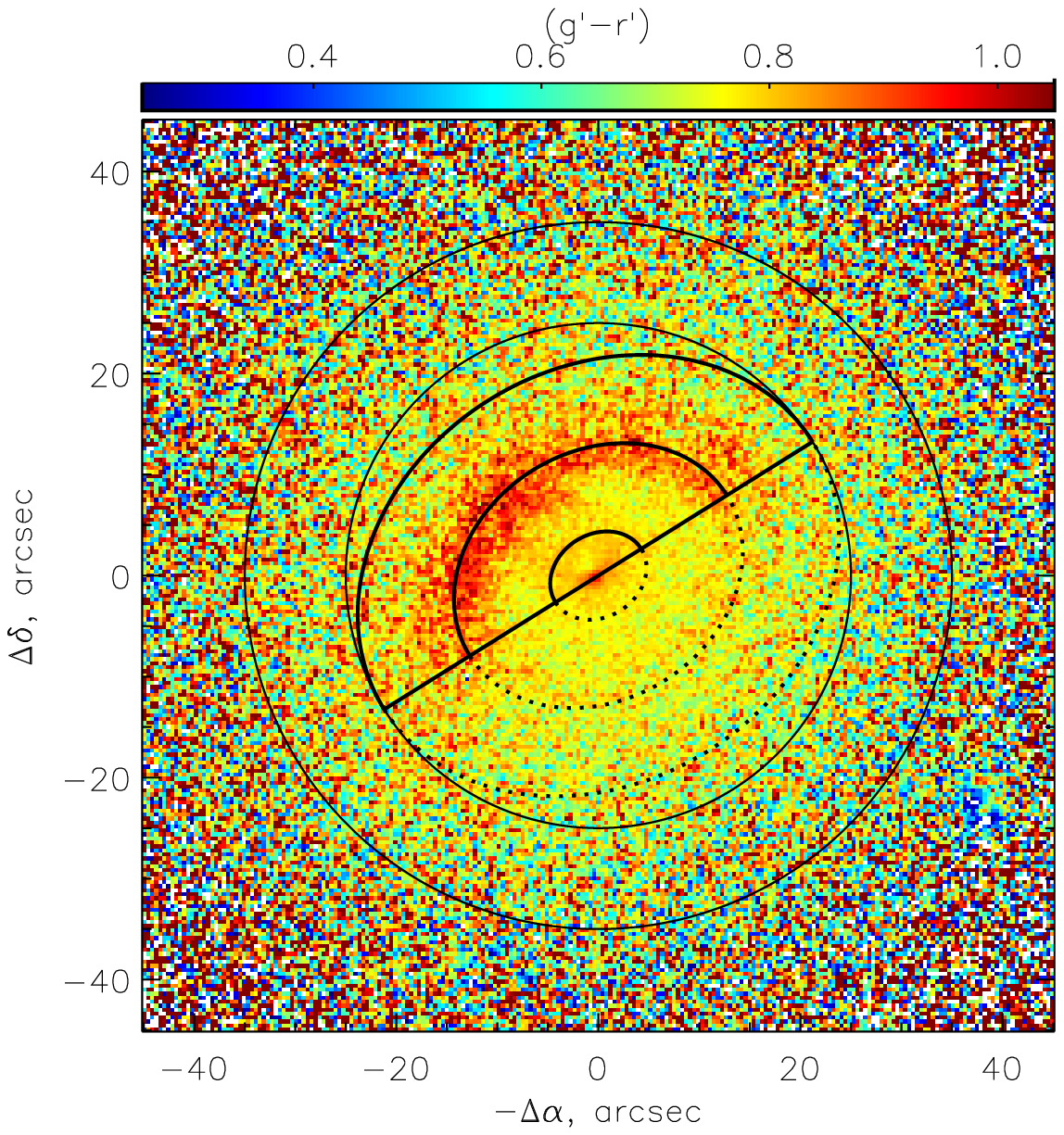}
\caption{The color map (g$'$-r$'$) of NGC 5631, from the SDSS data.
The $r'$-band isophotes (top) and schematic view of the both disks (bottom) are overlapped onto the color map. The derived orientation of the inner inclined
gaseous/dust disk is $i=35^\circ$, $PA=122^\circ$.}
\end{figure}

\acknowledgements
The 6m telescope is operated under the financial support of
the Ministry of Science and Education (registration number 01-43).
Our study of the galaxies with the multi-tiers disks, such as NGC~5631,
is supported by the grant of the
Russian Foundation for Basic Researches no. 07-02-00229.
A. V. M. acknowledges a grant from the President of the Russian Federation (MK1310.2007.2).
During the data analysis we have used the Lyon-Meudon Extragalactic Database (HYPERLEDA) supplied by the LEDA team at the CRAL-Observatoire de Lyon (France) and the NASA/IPAC
Extragalactic Database (NED) which is operated by the Jet Propulsion
Laboratory, California Institute of Technology, under contract with
the National Aeronautics and Space Administration.
This research is partly based on data obtained from the Isaak Newton Group Archive
which is maintained as part of the CASU Astronomical Data Centre at the Institute
of Astronomy, Cambridge,
on observations made with the NASA/ESA Hubble Space Telescope, obtained
from the data archive at the Space Telescope Science Institute, which is
operated by the Association of Universities for Research in Astronomy,
Inc., under NASA contract NAS 5-26555, and on SDSS data.
Funding for the Sloan Digital Sky Survey (SDSS) and SDSS-II has been provided by the Alfred P. Sloan Foundation, the Participating Institutions, the National Science Foundation, the U.S. Department of Energy, the National Aeronautics and Space Administration, the Japanese Monbukagakusho, and the Max Planck Society, and the Higher Education Funding Council for England. The SDSS Web site is http://www.sdss.org/.

\end{document}